\documentclass[prl,aps,twocolumn,showpacs,preprintnumbers,amsmath,amssymb]{revtex4}

\usepackage{graphicx}
\usepackage{dcolumn}

\begin{document}
\title{Superconductivity on the Localization Threshold\\
and Magnetic-Field-Tuned Superconductor-Insulator\\
Transition in TiN Films}

\author{T.~I.~Baturina$^{1,*}$,
        D.~R.~Islamov$^1$, J.~Bentner$^2$, C.~Strunk$^2$,
        M.~R.~Baklanov$^3$, A.~Satta$^3$}

\affiliation{$^1$Institute of Semiconductor Physics, 13 Lavrentjev Ave.,
630090 Novosibirsk, Russia\\
$^2$Institut f\"{u}r experimentelle und angewandte Physik,
Universit\"{a}t  Regensburg, D-93025 Regensburg, Germany\\
$^3$IMEC, Kapeldreef 75, B-3001 Leuven, Belgium\\
$^*$e-mail: tatbat@isp.nsc.ru
}

%
%
%


\begin{abstract} Temperature- and magnetic-field dependent measurements of the resistance
of ultrathin superconducting TiN films are presented. The analysis of the
temperature dependence of the zero field resistance indicates an underlying
insulating behavior, when the contribution of Aslamasov-Larkin fluctuations is taken
into account. 
This demonstrates the possibility of coexistence of the superconducting and 
insulating phases and of a direct transition from the one to the other. 
The scaling behavior of magnetic field data is in accordance with
a superconductor-insulator transition (SIT) 
driven by quantum phase fluctuations in two-dimensional superconductor. The
temperature dependence of the isomagnetic resistance data on the high-field side of
the SIT has been analyzed and the presence of an insulating phase is confirmed. A
transition from the insulating to a metallic phase is found at high magnetic fields,
where the zero-temperature asymptotic value of the resistance being equal to
$h/e^2$.
\end{abstract}


\maketitle

The interplay between superconductivity and localization is a phenomenon of
fundamental interest, and the question of the nature of superconductivity 
and its evolution in two-dimensional disordered systems 
and a perpendicular magnetic field
continues to receive a great deal of theoretical and experimental attention.
Two-dimensional systems are of special interest as two is the lower critical
dimensions for both localization and superconductivity. Two ground states are
expected to exist for bosons at $T=0$: a superconductor with long-range phase
coherence and an insulator in which the quantum mechanical correlated phase is
disjointed. 
The zero-temperature superconductor-insulator transition (SIT) is driven
purely by quantum fluctuations and is an example of a quantum phase
transition~\cite{Sondhi}. 
The superconducting phase is considered to be a condensate
of Cooper pairs with localized vortices, 
and the insulating phase is a condensate of
vortices with localized Cooper pairs. 
Between these two states there is the only metallic phase point, 
and this metal has a bosonic nature as well. The
theoretical description based on this assumption was suggested in~\cite{Fisher}. 
At finite temperatures, a quantum phase transition is influenced
by the thermal fluctuations, and according to the theory, 
(i)~the film resistance $R$ near the magnetic-field-induced SIT 
at low temperature $T$ in the vicinity of
the critical field $B_c$ is a function of one scaling variable
$\delta=(B-B_c)/T^{1/\nu z}$, with the critical exponents $\nu$ and $z$ being
constants of order of unity, and 
(ii)~at the transition point, the film resistance is of
the order $h/(2e)^2 \approx 6.5$~k$\Omega$ 
(the quantum resistance for Cooper pairs). 
Although much work has been done, and in many systems the scaling relations
hold~\cite{Hebard,Goldman99,Kapitulnik,Okuma,Destr,SITVFG}, the
magnetic-field-induced SIT in disordered films remains a controversial subject,
especially concerning the insulating phase and the bosonic 
conduction at $B>B_c$.
There is experimental evidence~\cite{Destr} that, despite 
the magnetoresistance being nonmonotonic, 
and in the magnetic fields above the critical one, the derivative of
resistance $dR/dT$ is negative, the phase can be insulating as well as metallic. The
behavior of the resistance in this region discussed 
in~\cite{Kapitulnik,Okuma} in terms of the magnetic-field-induced SIT 
(which is essentially {\it bosonic} in nature) 
can actually be explained on the basis of a {\it fermionic} approach, namely, in
the frames of the theory of the quantum corrections to the conductivity in
disordered metals. The possibility of such interpretation is shown in~\cite{SITvsQC}
based on the recent calculation of the quantum corrections due
to superconducting fluctuations~\cite{GalLar}. 
As a usual thermodynamic superconductor-normal metal transition, 
provided that the behavior of this metal is controlled, 
to a considerable degree, by the quantum corrections and a
superconductor-insulator transition may have very similar experimental
manifestations, some clear criteria are needed to enable one 
to tell which of the two underlies the behavior observed experimentally. 
Supposing the SIT to be the cause 
(e.g., the temperature dependence is found to be of an activated type), 
the question is then what effect 
the magnetic field may have on the bosonic insulator.

In this paper, we present the results of measurements and detailed analysis of
temperature and magnetic field dependence of the resistance of TiN films, 
devoting attention to a careful examination of the presence 
of the insulating phase and its
alteration on the high-field side of the SIT.
\begin{figure}[t]
 \includegraphics[width=3.1in]{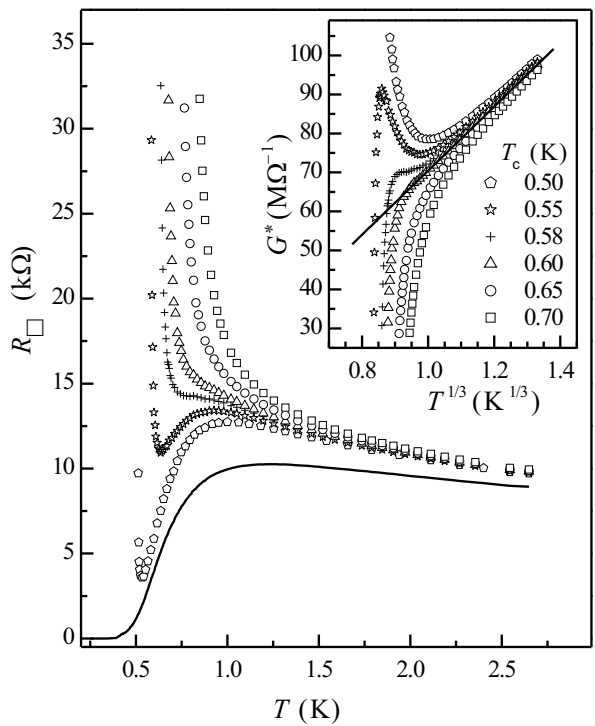}
 \caption{Temperature dependence of the resistance per square for sample 1 
  (solid line). Curves depicted by symbols correspond to $R^*(T_{\mathrm{c}})$ 
  (see text) and are obtained after
  subtraction of the Aslamasov-Larkin correction (\ref{AL_eq}) with different
  $T_{\mathrm{c}}$, which are listed in the inset. 
  These curves are presented as $G^*=1/R^*$ vs.~$T^{1/3}$ in the inset.}
 \label{fig1}
\end{figure}
\begin{figure}[t]
 \includegraphics[width=3.3in]{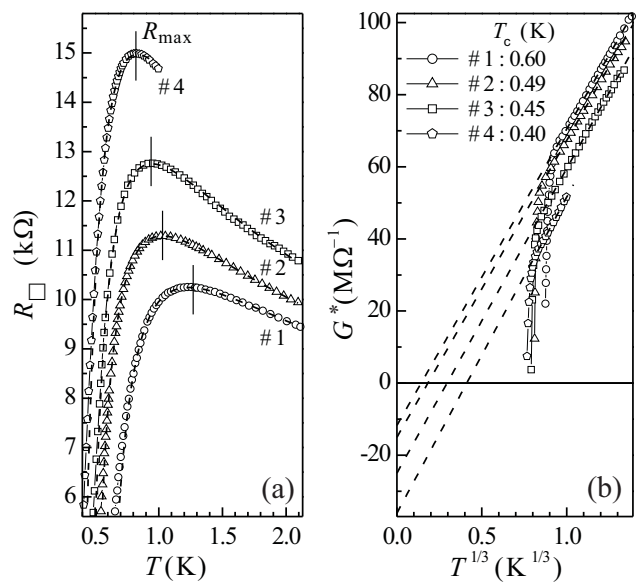}
 \caption{(a)~Temperature dependence of the resistance per square for
  four samples (symbols) and corresponding calculated curves $R_{AA+AL}$
  (dashed lines) at optimally fitted values of $a$ and $b$~(\ref{equAA}),
  and $T_{\mathrm{c}}$.
  (b)~$G^*=1/R^*$ vs. $T^{1/3}$ for the same samples at optimally fitted values of the critical
  temperature $T_{\mathrm{c}}$ listed in the figure.}
 \label{fig2}
\end{figure}

A TiN film with a thickness of 5~nm was formed on 100~nm of
SiO$_2$ grown on top of $<$100$>$ Si substrate by atomic layer
chemical vapor deposition at 350$^\circ$C \cite{ALCVD}.
Structural analysis shows that the formed TiN films are polycrystalline.
The films exhibit low surface roughness and consists of a dense packing
of the crystallities, with
a rather narrow distribution of size and an average size of
roughly $30$~nm. The samples for the transport measurements
were fabricated into Hall bridges using conventional
UV lithography and subsequent plasma etching.
Four terminal transport measurements were performed using standard low
frequency techniques. The resistance data were taken at a
measurement frequency of 10~Hz with an ac current of $0.04 \div 1$~nA.
The magnetic field was applied perpendicular to the film.

Four samples with the same thickness (5~nm) but different degrees of
disorder were studied in the present work. We begin by showing the temperature
dependence of the resistance $R(T,0)$ at zero magnetic field. $R(T,0)$ data are
presented in Fig.~\ref{fig1} for sample $1$ and in Fig.~\ref{fig2}
for all samples studied in this work. 
The resistance is a nonmonotonic function of the temperature, as is seen
more clearly in Fig.~\ref{fig2}a. 
With decrease of $T$, the increase of the resistance, which
is observed from $T=300$~K, is followed by a drop to the superconducting state. 
The transitions are significantly broadened. 
To explore reasons of such behavior and to
determine the main sample parameters, we apply an approach similar to the one used
previously in~\cite{GantGol}. 
As films under study are high resistive it should
be expected that the dependence $R(T)$ is strongly affected 
by the contribution of superconducting fluctuations 
(the Aslamazov-Larkin correction~\cite{ALAVR}) even at
the temperatures far from the transition temperature $T_{\mathrm{c}}$:
\begin{equation}
\Delta G_{AL} = \frac{e^2}{16\hbar}\left[\ln\left(
\frac{T}{T_{\mathrm{c}}}\right)\right]^{-1}.
\label{AL_eq}
\end{equation}
After extraction of this correction, with $T_{\mathrm{c}}$ being the only free
parameter, we obtain the temperature dependence $R^*(T_{\mathrm{c}})
=(1/R_\Box-\Delta G_{AL}(T_{\mathrm{c}}))^{-1}$, 
which is depicted by symbols in Fig.~\ref{fig1}. 
The curves $R^*(T_{\mathrm{c}})$ obtained with $T_{\mathrm{c}}<0.6$~K are
nonmonotonic, whereas the ones corresponding to $T_{\mathrm{c}}>0.6$~K 
give too strong growth of the resistance. 
The choice of $T_{\mathrm{c}}=0.6$~K is confirmed
by the further analysis of $R^*$, which is carried out 
for a start in terms of 3D ``bad'' metal in the vicinity of 
the metal-insulator transition (MIT) \cite{AAT13}.
In the critical region of the MIT, the behavior of the system is governed by
electron-electron interaction and the temperature dependence 
of the conductivity is
controlled by the only temperature-dependent scale $L_T=\sqrt{\hbar D/k_B T}$:
\begin{equation}
\sigma=\frac{e^2}{\hbar}\frac{1}{L_T}.
\label{sigmaAA}
\end{equation}
Using Einstein's relation $\sigma=e^2 D(\partial N / \partial\mu )$, 
the conductivity can be rewritten as
\begin{equation}
\sigma=\frac{e^2}{\hbar}\left(T \frac{\partial N}{\partial\mu}\right)^{1/3}.
\label{sigmaAA13}
\end{equation}
The representation of (\ref{sigmaAA13}) in the form of
\begin{equation}
G_{AA} (T)=a+bT^{1/3}
\label{equAA}
\end{equation}
is usually used to discriminate between a metal and an insulator
by means of determination of the sign of the parameter $a$ upon
extrapolation to $T=0$ (see, as an example,~\cite{ExpT13}).
A positive value of $a$ indicates the metallic state,
whereas a negative value of $a$ points to
activated conductance at lower temperatures.

The inset of Fig.~\ref{fig1} shows the conductance $G^*=1/R^*$ versus $T^{1/3}$. 
The dependence $G^*(T^{1/3})$ at $T_{\mathrm{c}}=0.6$~K 
is monotonic and the temperature, 
to which the linear character of $G^*$ is endured, is the lowest one. 
Determining in this way the values of $a$ and $b$ in~(\ref{equAA}) and
$T_{\mathrm{c}}$, we can fully describe the temperature dependence 
of the resistance film studied 
($R_{AA+AL}=[G_{AA}(a,b)+\Delta G_{AL}(T_{\mathrm{c}})]^{-1}$). 
The outcome of the above procedure applied to all samples under study 
is illustrated in Fig.~\ref{fig2}.
At $T>T_{\mathrm{c}}$, $R_{AA+AL}$ closely follows the
experimental data, justifying the validity of the above procedure 
and indicating that, with the deduction of
the direct contribution of the superconducting 
fluctuations, the conductance of the film is reasonably 
described by Eq.~(\ref{equAA}). 
It is interesting to note that the slope of the dependences $G^*(T^{1/3})$
plotted in Fig.~\ref{fig2}b (or the parameter $b$ in Eq.~(\ref{equAA})) 
is approximately the same for all samples
studied, whereas the values $a \equiv G^*(0)$ differ
significantly.

Additional information which can been extracted from the analysis
is the estimation of the parameters like the compressibility
\begin{equation}
\frac{\partial N}{\partial\mu}
=\frac{1}{k_{B}}\left(\frac{b\hbar}{e^2 d}\right)^3,
\label{dNdmu}
\end{equation}
the diffusion coefficient
\begin{equation}
D=\frac{k_B}{\hbar}\left(\frac{e^2 d}{b\hbar}\right)^2 T^{1/3},
\label{Dif}
\end{equation}
and $k_{F}l=3Dm/ \hbar$. Here, the value of $b$ is determined from the linear
approximation of $G^*$ versus $T^{1/3}$ (Eq.~\ref{equAA}) and $d$ is the thickness of the
film. The following numbers are obtained: $\partial N / \partial\mu = 3.6 \times
10^{21}$~eV$^{-1}$cm$^{-3}$, $D=0.29$~cm$^2/$s $(T/{\mathrm{K}})^{1/3}$, and
$k_{F}l=0.74(T/{\mathrm{K}})^{1/3}$, that is, $k_{F}l<1$ at $T<2.4$~K.

It should be noted that, in order to take into account the superconducting
fluctuations, we have used the Eq.~(\ref{AL_eq}), 
which is valid for the two-dimensional case. 
The condition under which a film may be considered two dimensional with
respect to superconducting fluctuations is
\begin{equation}
\ln \left(\frac{T}{T_{\mathrm{c}}}\right) \ll
\frac{\pi \hbar D}{8 k_B T_{\mathrm{c}} d^2}.
\label{2D_AL}
\end{equation}
Physically, this inequality denotes that the time of diffusive motion 
across the film is less than the fluctuation Cooper pair lifetime 
(or Ginzburg-Landau time)
\begin{equation}
\tau_{GL}^{-1}=\frac{8 k_B T}{\pi \hbar D}
\ln \left(\frac{T}{T_{\mathrm{c}}}\right).
\label{tauGL}
\end{equation}
Condition~(\ref{2D_AL}) can be rewritten 
as $d^2 \ll D\tau_{GL}=l_{GL}^2$. Using
the estimation of the diffusion coefficient, we find that $l_{GL}$ is 
larger than the film thickness at all temperatures under study.

One more relevant information can been received from the above analysis. 
Let us add~Eqs.~(\ref{sigmaAA13}) and~(\ref{AL_eq}) and 
rewrite total conductance as follows:
\begin{equation}
G=G_{AA}+\Delta G_{AL}=\frac{e^2}{\hbar}\frac{d}{L_T}
+\frac{e^2}{2 \pi \hbar}\frac{l_{GL}^2}{L_T^2}.
\label{sumG}
\end{equation}
Then, the temperature corresponding to the minimal value of $G$ is determined by
the condition
\begin{equation}
L_T=l_{GL} \left(\frac{12}{\pi ^2}\frac{l_{GL}}{d} \right)^{1/3},
\label{LTlGL}
\end{equation}
and, accordingly, the maximum of $R(T)$ 
($R_{\mathrm{max}}$ in Fig.~\ref{fig2}a) results from the condition $L_T \approx l_{GL}$.
Thus, the nonmonotonic temperature dependence of the resistance 
is the consequence of a competition of two length scales: 
$l_{GL}$ being responsible for Cooper pairing and
$L_T$ defining the electron-electron interaction. 
It should be stressed that the above estimations are rough, 
since the application of Eq.~(\ref{sigmaAA13}) implies 
that $a=0$ in~(\ref{equAA}). There is nothing like this
in our case, and, moreover, for the films under study, we get $a<0$.
It says that the underlying state is insulating.

\begin{figure}[t]
 \includegraphics[width=3.1in]{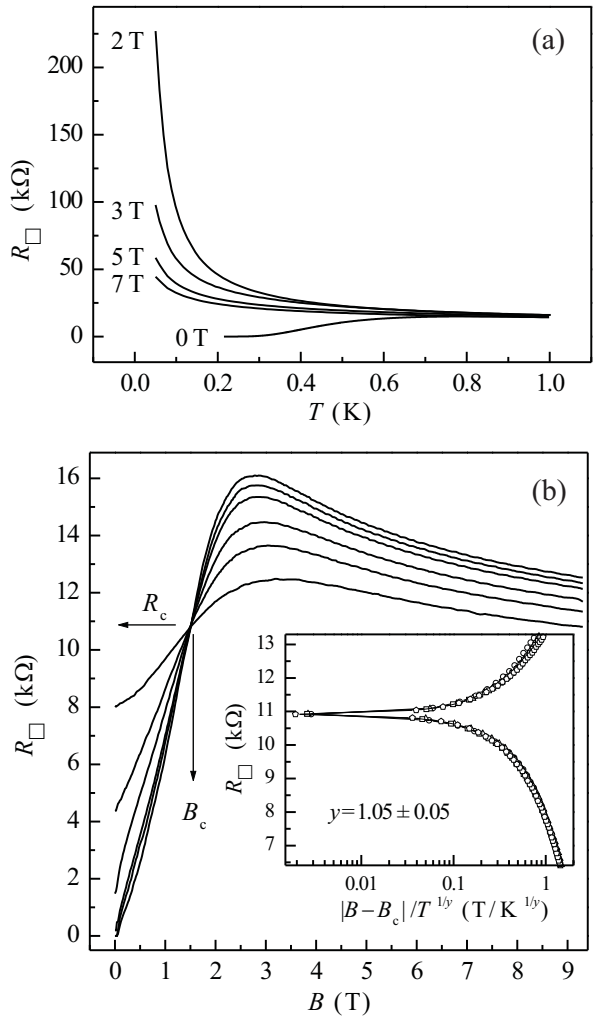}
 \caption{(a)~Dependences $R(T)$ at different $B$ of the sample $4$.
  (b)~Low-temperature isotherms of the sample $1$ in the $(B,R)$ plane.
  Different curves represent different temperatures: 0.35, 0.38,
  0.42, 0.51, 0.61, and 0.76 K. The point of intersection,
  $B_{c}=1.52$~T, is the critical magnetic field, and
  $R_{c}=10.9$~k$\Omega$ is the critical resistance. The inset shows
  a scaled plot of the same data with $y=\nu z=1.05\pm 0.05$.}
 \label{fig3}
\end{figure}
\begin{figure}[t]
 \includegraphics[width=3.3in]{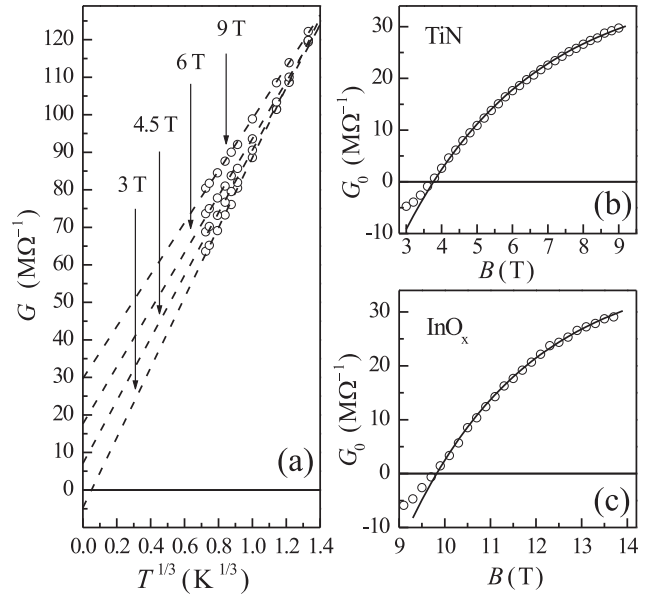}
 \caption{(a)~Conductance $G = 1/R$ vs. $T^{1/3}$ at different magnetic fields on the
  high-field side of the SIT for the sample~$1$. The magnetic field dependence of
  the zero temperature conductance determined from extrapolations in accordance with
  Eq.~(\ref{equAA}) are shown by symbols along with the dependences calculated from an
  empirical expression \ref{emp_Beq}: (b) sample~$1$ (this work) and (c)
  InO$_x$~\cite{SITVFG}.}
 \label{fig4}
\end{figure}

We now turn to the evolution of the resistance with temperature for various 
magnetic fields. Figure~\ref{fig3}a shows the isomagnetic temperature dependences 
of the resistance of sample $4$. 
Not too high magnetic field ($B < 2$~T) destroys the long-phase coherence
and reveals the underlying insulating state.
The high-field data appear to be more metallic in character: the magnetic
field results in a significant suppression 
of the insulating phase above 2~T. 
This is seen more clearly on a typical set of $R_\square(B)$ traces 
measured on sample~$1$ (see Fig.~\ref{fig3}b). 
The main feature of this graph is the presence of an intersection
point at $B_c, R_c$. Using the $B_c$, we plot the same data against the scaling
variable $|B-B_c|/T^{1/ \nu z}$ 
and adjust the product of the critical exponents $\nu z$ to obtain
the best scaling of the data (inset of Fig.~\ref{fig3}b).
However, such behavior, previously regarded as the main evidence of the existence of
SIT, is actually not incontestable proof of the presence of the insulating phase at
$B>B_c$ \cite{SITvsQC}. 
In order to ascertain the type of phase, we have analyzed the
temperature dependence of the high-field conductance 
at different magnetic fields.
In this regime, $G(T)$ is well described by Eq.~(\ref{equAA}), as indicated in
Fig.~\ref{fig4}a. As $G(0)$ is negative in fields higher than the critical field, we can
conclude that this phase is insulating. 
With further increase of $B$, the sign of
$G(0)$ changes, which points towards the transition to the metallic state. 
The extrapolation to $T=0$ allows us to determine not only the field of the
insulator-metal transition ($B_{IM}$) but also the magnetic field dependence 
of the zero-temperature conductance. 
The result of this procedure is presented in Fig.~\ref{fig4}b.
Analysis of the zero-temperature conductance at $B>B_{IM}$ 
reveals that it is well described by the empirical expression
\begin{equation}
G(T=0,B)=\frac{e^2}{h}\left(1-\exp \left[\frac{B_{IM}-B}{B^*}\right]\right)
\label{emp_Beq}
\end{equation}
shown by the solid line in Fig.~\ref{fig4}b. 
The magnetic-field-induced insulator-metal transition on the high field 
side of SIT was earlier observed in works of Gantmakher
{\it et al.} on InO$_x$~\cite{Destr,SITVFG}. 
Applying the same procedure to their
data (see left panel of Fig.~1 of Ref.~\cite{SITVFG}), 
we find that the dependence $G(T=0,B)$ is also well described 
by Eq.~(\ref{emp_Beq}) (see Fig.~\ref{fig4}c). 
The exponential dependence of $G(T=0,B)$ may result from a broad dispersion 
of the binding energies of localized Cooper pairs. 
The most important result is the zero-temperature asymptotic value 
of the resistance, which is equal to the quantum
resistance $R_Q=h/e^2$. 
The saturation of the low-temperature magnetoresistance to
the quantum resistance was demonstrated on beryllium films~\cite{Butko}.
Although authors of~\cite{Butko} consider that their films 
to be deep within the insulating phase, 
it is not improbable that they are superconducting at
lower temperature, i.e., that the correlated insulating phase 
consists of localized Cooper pairs. 
We believe that observed behavior, namely, the gradual approach of the
low-temperature magnetoresistance towards the quantum resistance, 
is a very general feature of a bosonic insulator on the high-field side 
of the magnetic-field-tuned SIT.

In conclusion, we have studied the temperature and magnetic field 
dependence of the resistance of TiN films. 
We have demonstrated that the nonmonotonic temperature
dependence of the resistance at zero magnetic field results from the 
concurrence of superconducting correlations and localization, 
with presumably the underlying insulating state having a bosonic nature.
The destruction of the long-range phase coherence by 
the magnetic field highlights this insulating state.
The further evolution of the system towards very high magnetic fields 
is in accordance with the breakup of such localized
Cooper pairs and drives the system eventually into a metallic regime. 
The saturation of the low-temperature magnetoresistance 
near the quantum resistance seems to be a
very general feature, occuring in several different materials. 
The nature of the insulating state and high-field metallic state 
requires further investigation.

We gratefully acknowledge discussions
with V.\,F.~Gantmakher, A.\,I.~Larkin and A.\,M.~Goldman.
We also thank V.\,F.~Gantmakher for
having given us access to his raw experimental data on InO$_x$.

This work was supported by the programs
``Superconductivity of Mesoscopic and Strongly Correlated Systems"
of the Russian Ministry of Industry, Science, and Technology,
``Low-Dimensional and Mesoscopic Condensed Systems" and 
``Quantum Macrophysics" of the Russian Academy of Science,
and by the Russian Foundation for Basic Research (grant no. 03-02-16368).

\vfill\eject

\end{document}